\documentclass[TDM,10pt]{iopart}

\usepackage{iopams}
\usepackage{graphicx}
\usepackage{color}
\usepackage{soul}
\usepackage[dvipsnames]{xcolor}

\newcommand{\mathsym}[1]{{}}

\def  \bsig    {\mbox{\boldmath$\sigma$}}

\begin{document}

\title[\footnotesize{Anomalous, spin, and valley Hall effects in graphene deposited on ferromagnetic substrates}]{Anomalous, spin, and valley Hall effects in graphene deposited on ferromagnetic substrates}

\author{A. Dyrda\l$^{1}$ and J.~Barna\'s$^{1,2}$}

\address{$^1$Faculty of Physics, Adam Mickiewicz University,
	ul. Umultowska 85, 61-614 Pozna\'n, Poland \\
	$^2$
	Institute of Molecular Physics, Polish Academy of Sciences,
	ul. M. Smoluchowskiego 17, 60-179 Pozna\'n, Poland
}
\ead{adyrdal@amu.edu.pl}
\vspace{10pt}
\begin{indented}
\item[]  \today 
\end{indented}

\begin{abstract}
	Spin, anomalous, and valley Hall effects in graphene-based hybrid structures are studied theoretically within the Green function formalism and linear response theory. Two different types of hybrid systems are considered in detail: (i) graphene/boron nitride/cobalt(nickel), and (ii) graphene/YIG. The main interest is focused on the proximity-induced exchange interaction between graphene and magnetic substrate and on the proximity-enhanced spin-orbit coupling. The proximity effects are shown to have a significant influence on the electronic and spin transport properties of graphene.  To find the spin, anomalous and valley Hall conductivities we employ certain effective Hamiltonians which have been proposed recently for the hybrid systems under considerations. Both anomalous and valley Hall conductivities have universal values when the Fermi level is inside the energy gap in the electronic spectrum.
\end{abstract}


%\submitto{\TDM}
%\maketitle
\ioptwocol
\begin{figure*}[h]
	\centering
	\includegraphics[width=0.95\textwidth]{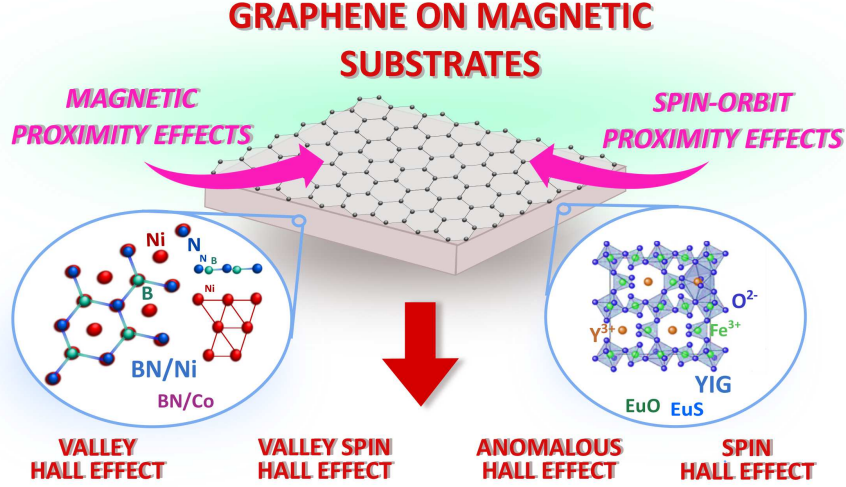}\\
	\caption{(Color online) Schematic of the system under consideration. Graphene is deposited either directly on a  magnetic substrate (YIG) or  is separated from the magnetic substrate (Co, Ni) by a few atomic planes of another hexagonal crystal (BN). The underlayer assures exchange coupling between the magnetic substrate and graphene  and also gives rise to the spin-orbit interaction of Rashba type. Owing to this, one may observe the Hall effects listed at the bottom of the figure. }
	\label{schematic}
\end{figure*}

\section{Introduction}
Graphene is a two-dimensional hexagonal lattice of carbon atoms. Electronic properties of pristine (or free standing) graphene have been extensively studied in recent years, mainly because of its unusual properties following from specific electronic states described by Dirac model~\cite{GeimNovoselov,Katsnelson_2007,Neto_RMP2009,Abergel,Roche_rev2017}. It has been shown that the electronic properties can be strongly modified when graphene is decorated (or functionalized)  with various adatoms or molecules attached to its surface or to edges in graphene stripes and nanoribbons~\cite{Gmitra2013,Balakrishnan2013,Irmer2015,Soriano_Fabian_Roche2015,Avsar2015,Zberecki2016}.

Other possibilities of a significant modification of graphene electronic and magnetic properties appear in hybrid systems based on graphene deposited on various substrates (e.g. on transition metal dichalcogenides or ferromagnetic thin films)~\cite{Avsar2014,Qiao2014,Wang2015,Gmitra2015,Gmitra2016,Yang2016,Zollner2016,vanWees2017,Hallal_Roche_2017}. Such systems  are currently of great interest both experimental and theoretical, mainly because of magnetic and spin-orbit proximity effects responsible for magnetic moment and enhanced spin-orbit interaction in the graphene layer. This, in turn, opens possibilities of spin-orbit driven phenomena in graphene-based hybrid structures at room temperatures~\cite{Balakrishnan2013,Avsar2014,Wang2015,vanWees2017,Mendes2015}. The high-temperature experimental realizations of anomalous and spin Hall effects as well as current-induced spin polarization (or Edelstein effect) make graphene-based structures active elements of future spintronics and spin-orbitronics devices -- together with other 2D crystals, semiconductor heterostructures, and junctions of oxide perovskites~\cite{Fert2016}.

Hexagonal two-dimensional crystals with their prominent examples such as graphene and transition metal dichalcogenides  with broken inversion symmetry are currently studied very intensively, especially in the context of so-called valleytronics and also valley-based optoelectronics~\cite{Yao2008,Xiao2012,Ezawa2013,Ezawa2014,Yamamoto2015,Ezawa2015,Song2015,Aivazian2015,Schaibley2016,Zhang2016}. An important property of such systems is the presence of two inequivalent ($K$ and $K'$) valleys in the corresponding electronic spectrum. Interestingly, it turned out that the valley degree of freedom can be controlled not only by circularly polarized light, but also with external magnetic and electric fields. Moreover, very promising for applications seems to be  coupling of the valley and spin degrees of freedom due to spin-orbit interaction~\cite{Heinz2014,Liu2015}. Owing to this one may expect, among others, a certain enhancement of the spin and valley polarization lifetimes and also manipulation of the spin degree of freedom by valley properties. This, in turn, allows to conceive a new generation of spintronic devices which are based on chargeless and nondissipative currents.

An important issue is the pure electrical generation and detection of valley and spin currents. This can be realized {\it via} the valley and spin Hall effects as well as their inverse counterparts. In systems with a net magnetization one can  also observe the anomalous Hall effect.  In high quality samples (free of defects and impurities), these effects may be determined by Berry curvature of the electronic bands and may reflect topological properties of the systems~\cite{Berry,Berry_rev_Xiao,Ezawa_book}. Thus,  detailed analysis of all the  Hall effects in graphene-based hybrid structures is crucial for their proper understanding.

In this paper we consider two kinds of hybrid structures: (i) graphene on a few atomic monolayers of boron nitride (BN) deposited on a ferromagnetic metal like Co or Ni, and (ii) graphene deposited directly on a ferromagnetic insulating substrate (YIG). In the former case the proximity-induced exchange interaction strongly depends on the number $n$ of atomic planes of BN, and disappears already for $n=4$ such atomic planes. In the latter case, in turn, the graphene layer is deposited  directly on the ferromagnetic substrate, so the exchange interaction is rather direct. Importantly, BN is a wide-gap semiconductor and therefore plays a role of energy barrier for low-energy electronic states in graphene.

Spin-orbit and exchange-interaction driven phenomena in graphene-based hybrid structures are studied within the linear response theory and Green function formalism. To describe these phenomena theoretically we make use of  the low-energy effective Hamiltonians, that have been derived recently from first-principle calculations (see e.g. \cite{Qiao2014,Wang2015,Zollner2016}). In particular, we calculate  the anomalous, spin and valley Hall conductivities. Apart from this, we also introduce the valley spin Hall effect. The anomalous and spin Hall effects occur due to spin-orbit coupling in the system subject to an external electric field (for review see~\cite{dyakonov71_she,hirsch,Engel2007,Nagaosa_rev2010,Sinova_rev2015}). In the case of  valley and valley spin Hall effects, the spin-orbit interaction is not required. Electrons have anomalous velocity component (normal to external electric field) which is oriented in opposite direction in the two  valleys (the corresponding Berry curvatures have opposite signs). As a consequence, electrons (or holes) from the two valleys are deflected towards opposite edges of the sample.
The above  effects  may play an important role in the graphene-based spintronics, as an effective source of spin currents and  spin-orbit torques~\cite{Dyrdal2015,Manchon2016}. These, in turn, may be responsible for spin dynamics and/or magnetic switching in the low-dimensional structures.

In section 2 we present a theoretical background and describe the model and theoretical method. In sections 3  we present our results on graphene/BN/Co and graphene/BN/Ni hybrid structures. Results for graphene/YIG hybrid system are presented and  discussed in section 4. Summary and final conclusions are in section 5.

\section{Theoretical background}

\subsection{Model}

We consider graphene either deposited directly on a magnetic substrate, or separated from the magnetic substrate by a few atomic layers of another two-dimensional crystal (e.g. BN), as shown schematically in Fig.\ref{schematic}. Influence of the substrate on magnetic and electronic properties of graphene  will be taken into account in terms of certain effective Hamiltonians which  have been obtained recently from results of {\it ab-initio} calculations~\cite{Qiao2014,Wang2015,Zollner2016}.
Because transport properties of graphene close to the charge neutrality point are determined mainly by electrons in the vicinity of Dirac points, we assume a minimal $p_{z}$ model that describes electronic and spin transport properties related to the low-energy electronic states of graphene and also takes into account the proximity-induced effects.

General low-energy Hamiltonian for both ($K$ and $K'$) Dirac points of the systems under considerations includes four terms~\cite{Zollner2016} ,
\begin{equation}
H^{\mathrm{K(K')}} = H_{0}^{\mathrm{K(K')}} + H_{\Delta}^{\mathrm{K(K')}} + H_{\scriptscriptstyle{\mathrm{EX}}}^{\mathrm{K(K')}} + H_{\mathrm{R}}^{\mathrm{K(K')}}.
\end{equation}
The first term of the above Hamiltonian describes electronic states of pristine graphene near the $K$ ($K'$) point~\cite{kane},
\begin{equation}
H_{0}^{\mathrm{K(K')}} = v (\pm k_{x} \sigma_{x} + k_{y} \sigma_{y}) s_{0},
\end{equation}
where $k_x$ and $k_y$ are the in-plane wavevector components, while $v = \hbar v_{F}$ with $v_{F}$ denoting the Fermi velocity.
Apart from this, we use the notation according to which  $\sigma_{0}$ and $\bsig$ are the unit matrix and the vector of Pauli matrices, $\bsig =(\sigma_x,\sigma_y, \sigma_z )$, acting in the pseudospin (sublattice) space, while  $s_{0}$ and $\mathbf{s}$ denote the unit matrix and vector of Pauli matrices, $\mathbf{s} = (s_{x}, s_{y}, s_{z})$, acting in the spin space.

The second term in Eq.(1) takes into account the fact that carbon atoms
from different sublattices (A and B) can feel generally different local potentials~\cite{Zollner2016,KochanIrmerFabian}. Such a dependence appears for instance when graphene is deposited on a 2D material with buckled or binary (like BN) hexagonal structure. This, in turn,  leads to the pseudospin symmetry breaking and gives rise to an orbital gap, $\Delta$, in the electronic spectrum,
\begin{equation}
H_{\Delta}^{\mathrm{K(K')}} = \Delta \sigma_{z} s_{0}.
\end{equation}

The third term in Hamiltonian (1) represents the proximity-induced  exchange interaction between graphene and magnetic substrate,  given explicitly by the formula~\cite{Zollner2016}
\begin{equation}
H_{\scriptscriptstyle{\mathrm{EX}}}^{\mathrm{K(K')}} = \frac{\lambda_{\scriptscriptstyle{\mathrm{EX}}}^{A}}{2} (\sigma_{z} - \sigma_{0}) s_{z} + \frac{\lambda_{\scriptscriptstyle{\mathrm{EX}}}^{B}}{2} (\sigma_{z} + \sigma_{0}) s_{z},
\end{equation}
where $\lambda_{\scriptscriptstyle{\mathrm{EX}}}^{A}$ and $\lambda_{\scriptscriptstyle{\mathrm{EX}}}^{B}$ are the exchange parameters corresponding to  the sublattices A and B, respectively. Note that in the special case of $\lambda_{\scriptscriptstyle{\mathrm{EX}}}^{A} = - \lambda_{\scriptscriptstyle{\mathrm{EX}}}^{B} = \lambda_{\scriptscriptstyle{\mathrm{EX}}}$ one obtains the exchange Hamiltonian in the form: $H_{\scriptscriptstyle{\mathrm{EX}}} = \lambda_{\scriptscriptstyle{\mathrm{EX}}} \sigma_{0} s_{z}$.

Finally, the last term in Hamiltonian (1) describes the spin-orbit interaction of Rashba type, that appears due to the space inversion symmetry breaking in the system. This interaction takes the following general form~\cite{kane}:
\begin{equation}
H_{\mathrm{R}}^{\mathrm{K(K')}} = \lambda_{\mathrm{R}} (\pm \sigma_{x} s_{y} - \sigma_{y} s_{x}),
\end{equation}
where $\lambda_{\mathrm{R}}$ is the Rashba parameter.
Note that the so-called intrinsic spin-orbit interaction in graphene is very small and therefore it is neglected in our consideration.

\subsection{Method}

Our key objective is to study the anomalous, spin, and valley Hall effects for graphene deposited on various substrates. Without loss of generality, we assume electric field along the axis $y$. The corresponding conductivities are determined by the contributions from both $K$ and $K'$ valleys as follows:
\begin{equation}
\sigma_{xy}^{\scriptscriptstyle{AHE}} = \sigma_{xy}^{K} + \sigma_{xy}^{K'}
\end{equation}
for the anomalous Hall effect (AHE),
\begin{equation}
\sigma_{xy}^{\scriptscriptstyle{VHE}} = \sigma_{xy}^{K} - \sigma_{xy}^{K'}
\end{equation}
for the valley Hall effect (VHE),
\begin{equation}
\sigma_{xy}^{\scriptscriptstyle{SHE}} = \sigma_{xy}^{s_{z}\,K} + \sigma_{xy}^{s_{z}\,K'}
\end{equation}
for the spin Hall effect (SHE), and
\begin{equation}
\sigma_{xy}^{\scriptscriptstyle{VSHE}} = \sigma_{xy}^{s_{z}\,K} - \sigma_{xy}^{s_{z}\,K'}
\end{equation}
for the valley spin Hall effect (VSHE).
Here, $\sigma_{xy}^{\nu}$ and $\sigma_{xy}^{s_{z}\,\nu}$ are contributions from the valley $\nu$ ($\nu =K,K'$) to the charge and spin conductivities, respectively.

Within the zero-temperature Green functions formalism and in the linear response with respect to a dynamical electric field of frequency $\omega$ (measured in energy units), one can  write the dynamical charge $\sigma_{xy}^{\nu}(\omega)$  and spin $\sigma_{xy}^{s_{z}, \nu}(\omega )$ conductivities  in the form,
\begin{eqnarray}
\sigma_{xy}^{\nu}(\omega) = \frac{e^{2} \hbar}{\omega} \int \frac{d\varepsilon}{2\pi} \int \frac{d^{2} \mathbf{k}}{(2\pi)^{2}} \nonumber \\
\hspace{1cm} \times \mathrm{Tr} \left\{\hat{v}_{x}^{\nu} G_{\mathbf{k}}^{\nu} (\varepsilon) \hat{v}_{y}^{\nu} G_{\mathbf{k}}^{\nu}(\varepsilon + \omega)  \right\},
\end{eqnarray}
\begin{eqnarray}
\sigma_{xy}^{s_{z}, \nu}(\omega) = \frac{e \hbar}{\omega} \int \frac{d\varepsilon}{2\pi} \int \frac{d^{2} \mathbf{k}}{(2\pi)^{2}} \nonumber \\
\hspace{1cm} \times \mathrm{Tr} \left\{\hat{j}_{x}^{s_{z}\,\nu} G_{\mathbf{k}}^{\nu} (\varepsilon) \hat{v}_{y}^{\nu} G_{\mathbf{k}}^{\nu}(\varepsilon + \omega)  \right\},
\end{eqnarray}
for $\nu =K, K' $.
In the above equations $\hat{v}_{x,y}^{\nu}$ denote  components of the velocity operator for the valley $\nu$,
$\hat{v}_{x,y}^{\nu} = \frac{1}{\hbar}\frac{\partial \hat{H}^{\nu}}{\partial k_{x,y}}$, while $\hat{j}_{x}^{s_{z}}$ is the relevant component of the spin current operator. Furthermore,  $G_{\mathbf{k}}^{\nu}(\varepsilon)$ stands for the causal Green function corresponding to the appropriate Hamiltonian $\hat{H}^{\nu}$, $G_{\mathbf{k}}^{\nu} = \{ [\varepsilon + \mu + i \delta\, {\rm sign}\,(\varepsilon)] - \hat{H}^{\nu}\}^{-1}$, where $\mu$ is the chemical potential and $\delta \rightarrow 0^{+}$ in the clean limit.

In the following we are interested in the dc-conductivities, so we take the limit $\omega\to 0$ in the above expressions.
To do this let us write
\begin{eqnarray}
\mathrm{Tr}\left\{ \hat{v}_{x}^{\nu} g_{\mathbf{k}}^{\nu}(\varepsilon+\omega) \hat{v}_{y}^{\nu} g_{\mathbf{k}}^{\nu} (\varepsilon)\right\}\nonumber \\
\hspace{1cm} = \mathcal{D}_{0}^{\nu}(\varepsilon, k, \phi) + \omega \mathcal{D}_{1}^{\nu}(\varepsilon, k, \phi) + ...
\end{eqnarray}
\begin{eqnarray}
\mathrm{Tr}\left\{ \hat{j}_{x}^{s_{z},\nu} g_{\mathbf{k}}^{\nu}(\varepsilon+\omega) \hat{v}_{y}^{\nu} g_{\mathbf{k}}^{\nu} (\varepsilon)\right\}\nonumber \\
\hspace{1cm} = \mathcal{D}_{0}^{s,\nu}(\varepsilon, k, \phi) + \omega \mathcal{D}^{s ,\nu}_{1}(\varepsilon, k, \phi) + ...,
\end{eqnarray}
where $g_{\mathbf{k}}^{\nu}$ stands for a nominator of the Green function,  $\phi$ is the angle between the wavevector $\mathbf{k}$ and the axis $y$, and the terms of higher order in $\omega$ have been omitted as their contribution vanishes in the limit of $\omega \to 0$.
Upon calculating the trace one finds $\mathcal{D}_{0}^{\nu}(\varepsilon, k, \phi)=0$ and $\mathcal{D}_{0}^{s,\nu}(\varepsilon, k, \phi)=0$. Thus, in the limit of $\omega \rightarrow 0$  the expressions (10) and (11) take  the  form
\begin{equation}
\sigma_{xy}^{\nu} = \frac{e^{2} \hbar}{(2\pi)^{3}} \int d\varepsilon \int dk\, k \mathcal{F}^{\nu}(\varepsilon ,k),
\end{equation}
\begin{equation}
\sigma_{xy}^{s_{z},\nu} = \frac{e \hbar}{(2\pi)^{3}} \int d\varepsilon \int dk\, k \mathcal{F}^{s,\nu}(\varepsilon ,k),
\end{equation}
where the functions $\mathcal{F}^{\nu}(\varepsilon ,k)$ and $\mathcal{F}^{s,\nu}(\varepsilon ,k)$ are defined as
\begin{equation}
\mathcal{F}^{\nu}(\varepsilon ,k) = \frac{\mathcal{I}^{\nu}(\varepsilon, k)}{\prod_{l = 1}^{4}[\varepsilon + \mu - E_{l} + i\delta{\mathrm{sgn}}(\varepsilon)]^{2}},
\end{equation}
\begin{equation}
\mathcal{F}^{s,\nu}(\varepsilon ,k) = \frac{\mathcal{I}^{s,\nu}(\varepsilon, k)}{\prod_{l = 1}^{4}[\varepsilon + \mu - E_{l} + i\delta{\mathrm{sgn}}(\varepsilon)]^{2}}.
\end{equation}
Here, $E_l$ ($l=1-4$) denote the four eigenmodes of the relevant Hamiltonian, and we introduced the following notation:
\begin{equation}
\mathcal{I}^{\nu}(\varepsilon, k) = \int d\phi \, \mathcal{D}_{1}^{\nu}(\varepsilon, k, \phi),
\end{equation}
\begin{equation}
\mathcal{I}^{s,\nu}(\varepsilon, k) = \int d\phi \,\mathcal{D}_{1}^{s,\nu}(\varepsilon, k, \phi).
\end{equation}

The integration over $\varepsilon$ in Eqs (14) and (15) can be performed in terms of the theorem of residues. As a result one finds
\begin{equation}
\sigma_{xy}^{\nu} = \frac{e^{2} \hbar}{(2 \pi)^{3}} \sum_{l = 1}^{4} \int dk\, k R_{l}^{\nu}f(E_{l}),
\end{equation}
\begin{equation}
\sigma_{xy}^{s_z,\nu} = \frac{e \hbar}{(2 \pi)^{3}} \sum_{l = 1}^{4} \int dk\, k R_{l}^{s, \nu}f(E_{l}),
\end{equation}
for $\nu =K, K'$.
Here,  $f(E)$ is the Fermi distribution function, while $R_{l}^{\nu}$ and $R_{l}^{s,\nu}$ denote the residua (multiplied by the factor $2\pi i$) of the functions $\mathcal{F}^{\nu}(\varepsilon ,k)$ and $\mathcal{F}^{s,\nu}(\varepsilon ,k)$, respectively,  taken at $\varepsilon = E_{l} - \mu$.

Since we consider here only intrinsic (topological) contributions to the anomalous and valley Hall effects, one can express Eq.(20) alternatively in terms of the Berry curvature of electronic bands corresponding to the valley $\nu$,
\begin{eqnarray}
\sigma_{xy}^{\nu}
= \frac{e^{2}}{\hbar} \sum_{l = 1}^{4} \int \frac{dk k}{(2\pi)^{2}} \bar{\Omega}_{l}^{\nu} f(E_{l}) \nonumber \\
= \frac{e^{2}}{\hbar} \sum_{l = 1}^{4} \int \frac{d^{2} \mathbf{k}}{(2\pi)^{2}} \Omega_{l}^{\nu} f(E_{l}),
\end{eqnarray}
where $\Omega_{l}^{\nu}$ is the $z$ component of the Berry curvature for the $l$-th subband, calculated in the vicinity of the point $\nu$,
while  $\bar{\Omega}_{l}^{\nu}$ is the Berry curvature integrated over the angle $\phi$,  $\bar{\Omega}_{l}^{\nu} = \int d\phi \,\Omega_{l}^{\nu}$.
Thus, the Berry curvature can be related to the residua $R_l^\nu$ as $\bar{\Omega}_{l}^{\nu}=2\pi\hbar^2R_l^\nu$.The correspondence between Kubo formulation and the approach based on topological invariants has been shown by Tholuess et al.~\cite{TKNN,Thouless_1983} and then it was widely discussed in the literature (see review papers \cite{Berry_rev_Xiao,Berry_rev_Niu}). Therefore, we only comment here that in the case of AHE and VHE, the conductivity may be nonzero even if the energy bands are described by the zero Chern number (Berry phase). This is because the local Berry curvature may be nonzero and can give rise to the  anomalous or valley Hall conductivity. This is the case that we consider in this paper.

Equations (20)-(22) are our general formulas which can be used to determine all the four Hall conductivities. These formulas  will be applied in the following to specific hybrid systems under consideration.

\section{Graphene/BN(n)/C\lowercase{o}(N\lowercase{i})}

Consider first graphene on a few ($n$) atomic planes of hexagonal BN which is deposited on ferromagnetic Co or Ni. Since BN has a wide energy gap, it can be considered as an insulating barrier.  Thus, the influence of Co (or Ni) on transport properties of graphene in the low-energy region is determined mainly by exchange interaction between graphene end Co (Ni) through the BN layer. It has been concluded from {\it ab-initio} calculations that  Rashba interaction in graphene/BN/Co(Ni)  hybrid system is much smaller than the exchange term and can be ruled out~\cite{Zollner2016}. Therefore, we consider the limit of vanishing Rashba interaction.
The relevant parameters extracted  from {\it ab-initio} calculations for graphene/BN/Co(Ni) systems by Zollner {\it et al}~\cite{Zollner2016} are given in Table 1. These parameters will be used below in our model calculations.
\begin{table}[h]
	\caption{\label{TabPar} Parameters describing graphene(Gr)-based hybrid systems under considerations, taken from
Ref.~\cite{Zollner2016}.}
	\footnotesize
\centering
	\begin{tabular}{@{}lllll}
		\br
		&n& $\Delta\, [meV]$ & $\lambda_{\scriptscriptstyle{\mathrm{EX}}}^{A}\, [meV]$& $\lambda_{\scriptscriptstyle{\mathrm{EX}}}^{B}\, [meV]$\\
		\mr
		Gr/BN/Co &1 & 19.25 & -3.14 & 8.59\\
		& 2 & 36.44 & 0.097 & -9.81\\
		& 3 & 38.96 & -0.005 & 0.018\\
		\mr
		Gr/BN/Ni & 1 & 22.86 & -1.40 & 7.78\\
		& 2 & 42.04 & 0.068 & -3.38 \\
		&  3 & 40.57 & -0.005 & 0.017 \\
		\br
	\end{tabular}\\
%	$^{a}$.....
	
\end{table}
\normalsize

\begin{figure*}[t]
	\centering
	\includegraphics[width=0.75\textwidth]{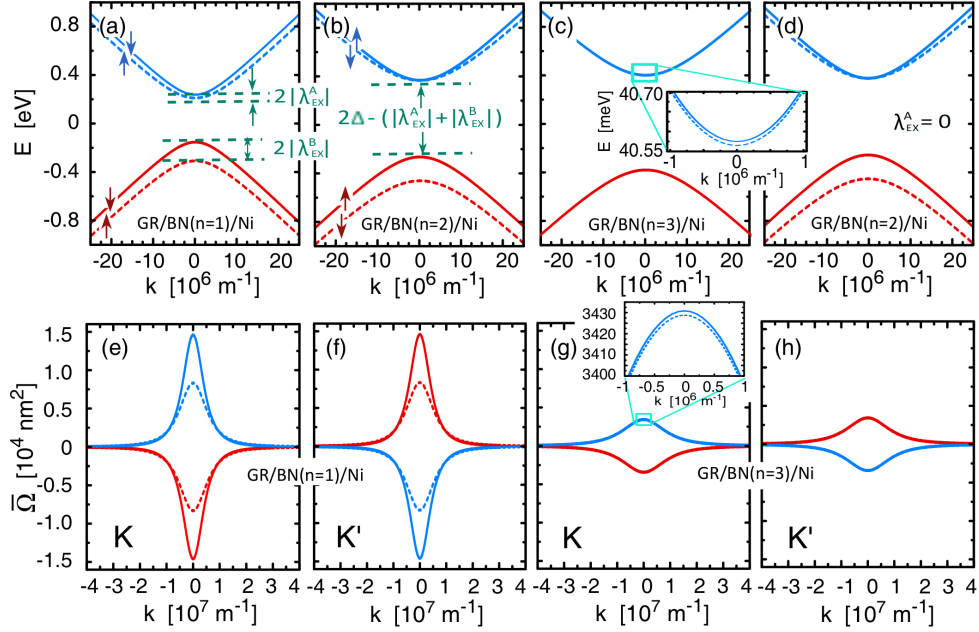}
	\caption{(Color online) Energy dispersion curves around the K point for $n=1$ (a), $n=2$ (b), and $n=3$ (c) atomic planes of BN and for the parameters presented in Table 1. Dispersion curves in a special case of $\lambda_{\scriptscriptstyle{\mathrm{EX}}}^A =0$ are shown in (d). Bottom panel shows the Berry curvature $\bar{\Omega}$ of electronic bands for $n=1$ (e,f) and $n=3$ (g,h) atomic planes of BN in the vicinity of both $K$ (e,g) and $K'$ (f,h) Dirac points. The arrows indicate spin polarization of the electronic bands.}
	\label{DISP_G/hBN/mgt}
\end{figure*}

When Rashba coupling disappears, Hamiltonian for the graphene/BN/Co(Ni) hybrid system can be reduced to the form
\begin{equation}
H^{\mathrm{K(K')}} = H_{0}^{\mathrm{K(K')}} + H_{\Delta}^{\mathrm{K(K')}} + H_{\scriptscriptstyle{\mathrm{EX}}}^{\mathrm{K(K')}}.
\end{equation}
The corresponding dispersion relations for the $K$ valley are shown in Fig.\ref{DISP_G/hBN/mgt} (top panel) for $n=1$, $n=2$ and $n=3$ monolayers of BN. Splitting of the conduction and valence bands due to exchange interaction, clearly seen for $n=1$ (Fig.\ref{DISP_G/hBN/mgt}a), becomes reduced for $n=2$ (Fig.\ref{DISP_G/hBN/mgt}b) and is negligible for $n=3$ (Fig.\ref{DISP_G/hBN/mgt}c). This is a consequence of reduced exchange interaction when the number of atomic planes of BN increases. Note, splitting of the valence band is remarkably larger than that of the conduction band.
Another interesting property of the spectrum is a relatively wide energy gap due to inversion symmetry breaking. This orbital gap is a consequence of the presence of BN layers, and its width increases with increasing number $n$ of BN monolayers. Interestingly, the gap is much wider than that in the free standing graphene, where it is negligible due to  a very small intrinsic spin-orbit interaction.

Berry curvature integrated over the angle $\phi$  is shown in  Fig.\ref{DISP_G/hBN/mgt} (bottom panel) for $n=1$, $n=3$, and for both $K$ and $K'$ valleys. This figure clearly shows that the curvature of electronic bands in the $K$ valley is opposite to the corresponding curvature in the $K'$ valley.
As a result one finds  $\sigma_{xy}^{K'} = - \sigma_{xy}^{K}$ in the case under consideration, i.e. contributions to the anomalous Hall conductivity from individual valleys are not zero, but they have opposite signs and cancel each other. Therefore, the anomalous Hall effect vanishes, $\sigma_{xy}^{\scriptscriptstyle{AHE}} = 0$. This is rather clear as there is no spin-orbit interaction. Similarly, also the SHE vanishes due to the lack of spin-orbit coupling.
However, the VHE effect remains nonzero, $\sigma_{xy}^{\scriptscriptstyle{VHE}} = 2 \sigma_{xy}^{K} $, and from similar reasons also the VSHE is nonzero.
%Thus, based on  Eq.(9) we get: ...

\subsection{Valley Hall effect}

Due to the opposite Berry curvature of the electronic bands in the $K$ and $K'$ valleys, electrons from both valleys are deflected to opposite edges, giving rise to a nonzero valley Hall conductivity (see Eq.~(7).
Simple analytical results can be derived in a specific case of $\lambda_{\scriptscriptstyle{EX}}^{A} = 0$. The corresponding dispersion curves around the $K$ point are shown in Fig.\ref{DISP_G/hBN/mgt}d. Note, the conduction band is then degenerate at $k=0$. Detailed analytical calculations show that the valley Hall conductivity depends on the Fermi level $\mu$, and bearing in mind that $\Delta > |\lambda_{\scriptscriptstyle{EX}}^{B}|$ this dependence  can be written as follows: \\
\vspace{0.5cm}\\
(i)  $-\Delta + |\lambda_{\scriptscriptstyle{EX}}^{B}| < \mu < \Delta$ (Fermi level inside the gap):
\begin{equation}
\sigma_{xy}^{\scriptscriptstyle{VHE}} = -2 \frac{e^{2}}{\hbar},
\end{equation}
i.e. the valley Hall conductivity is quantized. \\
(ii) $\mu > \Delta$ (Fermi level inside the conduction bands),
or $\mu < - (\Delta + |\lambda_{\scriptscriptstyle{EX}}^{B}|)$ (Fermi level inside the valence bands):
\begin{equation}
\sigma_{xy}^{\scriptscriptstyle{VHE}} = - \left(\frac{2\Delta + \lambda_{\scriptscriptstyle{EX}}^{B}}{|2 \mu + \lambda_{\scriptscriptstyle{EX}}^{B}|} + \frac{2 \Delta - \lambda_{\scriptscriptstyle{EX}}^{B}}{|2 \mu - \lambda_{\scriptscriptstyle{EX}}^{B}|} \right) \frac{e^{2}}{h}.
\end{equation}
(iii)  $-(\Delta + |\lambda_{\scriptscriptstyle{EX}}^{B}|) < \mu < -\Delta + |\lambda_{\scriptscriptstyle{EX}}^{B}|$: \begin{equation}
\sigma_{xy}^{\scriptscriptstyle{VHE}} = - \left( 1 + \frac{2 \Delta - |\lambda_{\scriptscriptstyle{EX}}^{B}|}{|2 \mu - |\lambda_{\scriptscriptstyle{EX}}^{B}||}\right)\frac{e^{2}}{h}.
\end{equation}

\begin{figure*}
	\centering
	\includegraphics[width=0.7\textwidth]{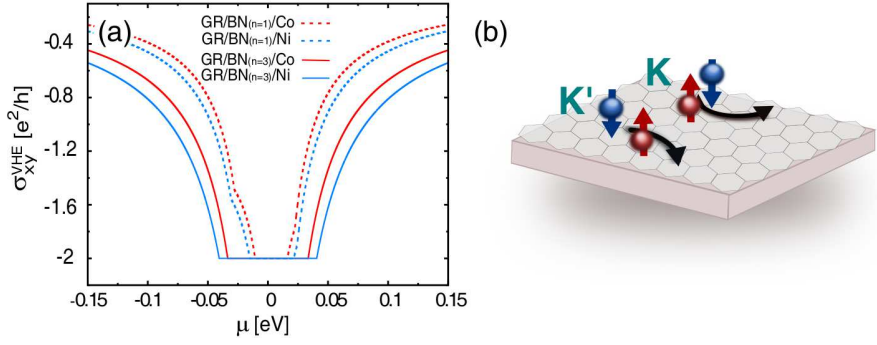}\\
	\caption{(Color online) (a) Valley Hall conductivity as a function of the chemical potential $\mu$ for graphene/BN/Co and graphene/BN/Ni systems with $n=1$ and $n=3$ atomic planes of BN. (b) Schematic presentation of the VHE: electrons in the $K$ and $K'$ valleys are deflected in opposite orientations normal to external electric field.
	\label{VHE}}
\end{figure*}

In a general situation, $\lambda_{\scriptscriptstyle{EX}}^{A} \ne 0$, the valley Hall conductivity was calculated numerically and is shown in  Fig.\ref{VHE} as a function of the chemical potential $\mu$. The  valley conductivity is quantized for the Fermi level in the gap, where
$\sigma_{xy}^{\scriptscriptstyle{VHE}} = -2 \frac{e^{2}}{\hbar}$. The absolute value of the conductivity for $\mu$ outside the gap is reduced with increasing $|\mu |$. The kinks appear at the points where the Fermi level crosses edges of the conduction or valence bands, and appear in the presence of exchange splitting of the bands. Note, such a splitting disappears for $n=3$ monolayers of BN, where the exchange interaction is vanishingly small. The kinks for positive $\mu$ are less pronounced as the exchange-induced splitting of the conduction band is remarkably smaller.

\subsection{Valley spin Hall effect}
\begin{figure*}
	\centering
	\includegraphics[width=\textwidth]{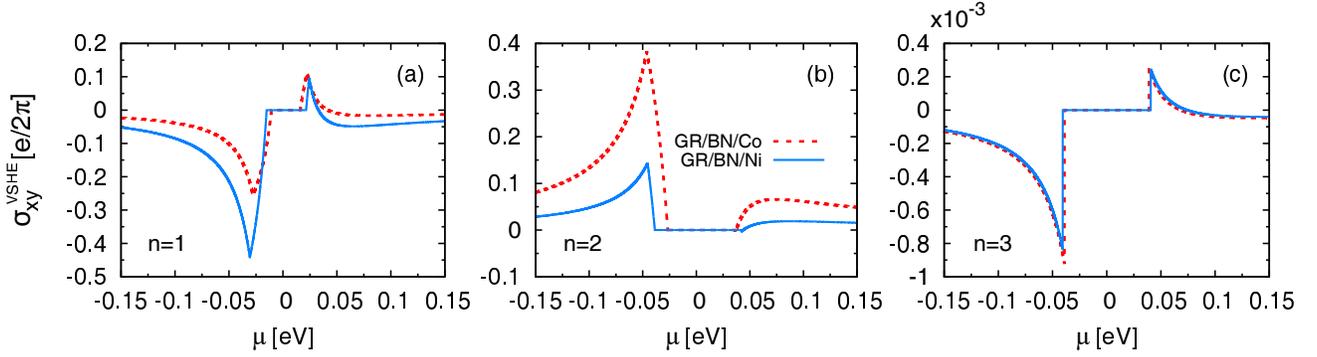}\\
	\caption{(Color online)  Valley spin Hall conductivity as a function of the chemical potential $\mu$ for graphene/BN/Co (dashed lines) and graphene/BN/Ni (solid lines) systems with $n=1$ (a), n=2 (b), and $n=3$ (c) atomic planes of BN. Due to small exchange coupling, the valley spin Hall conductivity for $n=3$ is by three orders of magnitude smaller. The sign of conductivity for $n=2$ is reversed due to reversed sign of the exchange parameter.}
	\label{VSHE}
\end{figure*}

As already mentioned above, the spin Hall effect vanishes in graphene/BN/Co (Ni) systems due to vanishingly small Rashba interaction.  Strictly speaking, contributions to the spin Hall conductivity from individual valleys are nonzero, however they cancel each other as the spin currents associated with the $K$ and $K'$ valleys are opposite. Thus, similarly to the valley Hall effect,  one can define the valley spin Hall effect as the difference of spin currents from the $K$ and $K'$ valleys, see Eq.(9). This quantity is generally nonzero, and indicates that the net spins from the  $K$ and $K'$  valleys are deflected to the opposite edges. As in the previous section we analyse the full model with a finite parameter $\lambda_{\scriptscriptstyle{EX}}^{A}$, as well as the limit $\lambda_{\scriptscriptstyle{EX}}^{A} = 0$. For vanishing $\lambda_{\scriptscriptstyle{EX}}^{A}$ it is possible to find analytical solutions for the valley spin Hall conductivity. \\
(i)  $\mu > \Delta$:
\begin{equation}
\sigma_{xy}^{\scriptscriptstyle{VSHE}} = \frac{e}{2\pi} \left( \frac{2\Delta - \lambda_{\scriptscriptstyle{EX}}^{B}}{|2\mu - \lambda_{\scriptscriptstyle{EX}}^{B}|} - \frac{2 \Delta + \lambda_{\scriptscriptstyle{EX}}^{B}}{|2\mu + \lambda_{\scriptscriptstyle{EX}}^{B}|}\right).
\end{equation}
(ii) $-\Delta + |\lambda_{\scriptscriptstyle{EX}}^{B}| < \mu < \Delta$ (Fermi level is in the gap):\\
\begin{equation}
\sigma_{xy}^{\scriptscriptstyle{VSHE}} = 0,
\end{equation}
i.e. the valley spin Hall conductivity vanishes. \\
(iii)  $- \Delta - |\lambda_{\scriptscriptstyle{EX}}^{B}| < \mu < - \Delta + |\lambda_{\scriptscriptstyle{EX}}^{B}|$:
\begin{equation}
\sigma_{xy}^{\scriptscriptstyle{VSHE}} = - \frac{e}{2\pi} \left( 1 - \frac{2 \Delta - |\lambda_{\scriptscriptstyle{EX}}^{B}|}{|2\mu - |\lambda_{\scriptscriptstyle{EX}}^{B}||}\right).
\end{equation}
(iv) $\mu < - \Delta - |\lambda_{\scriptscriptstyle{EX}}^{B}|$:
\begin{equation}
\sigma_{xy}^{\scriptscriptstyle{VSHE}} = - \frac{e}{2\pi} \left( \frac{2 \Delta + \lambda_{\scriptscriptstyle{EX}}^{B}}{|2\mu + \lambda_{\scriptscriptstyle{EX}}^{B}|} - \frac{2 \Delta - \lambda_{\scriptscriptstyle{EX}}^{B}}{|2\mu - \lambda_{\scriptscriptstyle{EX}}^{B}|}\right).
\end{equation}

Numerical results on the valley spin Hall conductivity are presented in Fig.\ref{VSHE} for the general situation, $\lambda_{\scriptscriptstyle{EX}}^{A} \ne 0$, and for $n=1$, \textcolor{PineGreen}{$n=2$} and $n=3$.
The valley spin Hall conductivity vanishes for the Fermi level in the gap. To understand this we note first that the  exchange-splitting of  conduction (and also valence) bands is the same in the $K$ and $K'$ valleys. Since the two valence subbands in an individual valley correspond to opposite spin orientations, their contributions to the spin current  exactly cancel each other when the Fermi level is in the energy gap. A nonzero spin current appears then when the Fermi level crosses the bottom edge of the lower conduction  subband or top edge of the higher valence subband. When $|\mu |$ groves further, the valley spin Hall conductivity decreases due to compensating contribution from the second conduction (valence) subband.

Note, the valley spin Hall conductivity for $n=2$ atomic planes of BN has reversed sign in major part of $\mu$ due to reversed sign of the exchange parameter in comparison to that for $n=1$. Apart from this, the valley spin Hall conductivity for $n=3$ is roughly three orders of magnitude smaller than for $n=2$. This is due to a very small exchange coupling parameter for $n=3$.

\section{Graphene on a magnetic insulating substrate}

\begin{figure*}[t]
	\centering
	\includegraphics[width=0.85\textwidth]{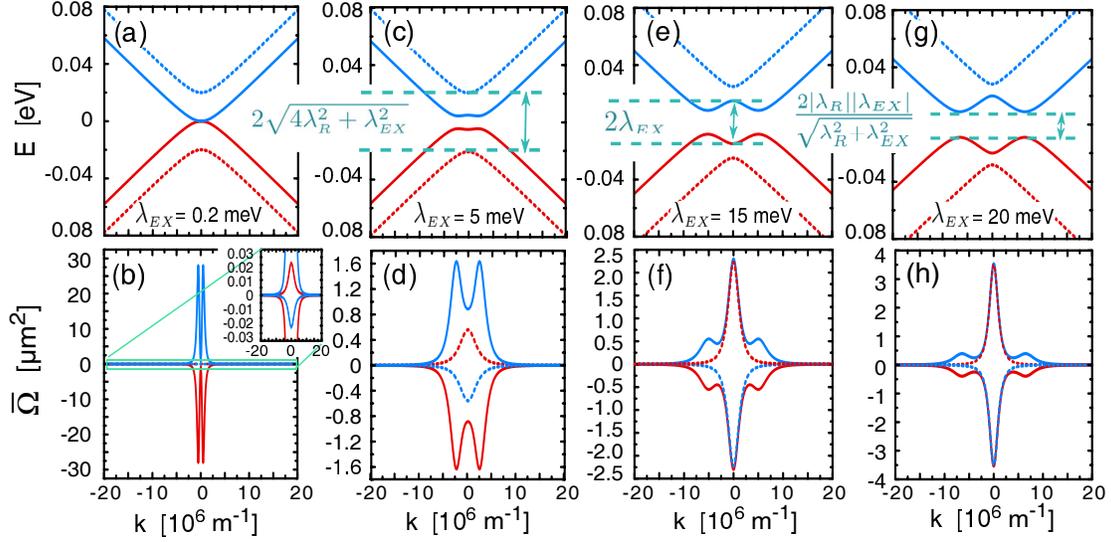}\\
	\caption{(Color online)  (a,c,e,g) Energy dispersion curves around the $K$ Dirac point in the graphene/YIG system for a constant Rashba parameter and exchange parameter as indicated. (b,d,f,h) Berry curvature integrated over the angle $\phi$ corresponding to the bands shown in (a,c,e,g), respectively. }
	\label{DISP_G/YIG}
\end{figure*}
Now we consider graphene deposited directly on a magnetic insulating substrate. An important example of such a hybrid system is graphene deposited on YIG, where large anomalous Hall effect at room temperature has been measured recently~\cite{Wang2015,vanWees2017}. In this particular case the third term of Hamiltonian (1), corresponding to the orbital gap, is absent. However, the coexistence of proximity-induced exchange field and Rashba spin-orbit coupling is essential. Therefore, Hamiltonian (1) for the graphene/YIG system can be reduced to the following one:
\begin{equation}
\label{H_GonYIG}
H^{\mathrm{K(K')}} = H_{0}^{\mathrm{K(K')}} +  H_{\scriptscriptstyle{\mathrm{EX}}}^{\mathrm{K(K')}} + H_{\mathrm{R}}^{\mathrm{K(K')}}.
\end{equation}
Moreover, one may assume $\lambda_{\scriptscriptstyle{\mathrm{EX}}}^{A} = - \lambda_{\scriptscriptstyle{\mathrm{EX}}}^{B} = \lambda_{\scriptscriptstyle{\mathrm{EX}}}$ in this particular case, so  the relevant exchange Hamiltonian reads
\begin{equation}
H_{\scriptscriptstyle{\mathrm{EX}}}^{\mathrm{K(K')}} = \lambda_{\scriptscriptstyle{\mathrm{EX}}} \sigma_{0} s_{z}.
\end{equation}

Figure \ref{DISP_G/YIG} presents the energy dispersion curves for the graphene/YIG structure (top panel). The Rashba coupling was assumed there constant while the exchange parameter  was changed (as indicated)  from weak to strong coupling limit. Interestingly, when the exchange coupling is small, there is no energy gap in the spectrum -- the gap is created when the exchange interaction is sufficiently strong. Apart from this, minima (maxima) of the conduction (valence) bands are shifted away from the Dirac points. The bottom panel in Fig.~\ref{DISP_G/YIG} shows the Berry curvature corresponding to the bands displayed in the top panel. The Berry curvature for the $K'$ point (not shown) is the same as that for the $K$ point. Due to to this, both VHE and VSHE are absent. However,  AHE and SHE conductivities do not vanish due to Rashba spin-orbit coupling, and both can be found following the approach described in section 2.

\subsection{Spin Hall effect}

To find the spin Hall conductivity we make use of Eq.(21). The corresponding residua can be easily evaluated and are given by  the expressions
\begin{equation}
R_{1(3)}^{K,s} = \pi^{2}\frac{\lambda_{R}^{2} v^{2} (2 (\lambda_{R}^{2} + \lambda_{\scriptscriptstyle{EX}}^{2}) + v^{2} k^{2})}{(\lambda_{R}^{4} + v^{2} k^{2} (\lambda_{R}^{2} + \lambda_{\scriptscriptstyle{EX}}^{2}))^{3/2}},
\end{equation}
\begin{equation}
R_{2,(4)}^{K,s} = -\pi^{2}\frac{\lambda_{R}^{2} v^{2} (2 (\lambda_{R}^{2} + \lambda_{\scriptscriptstyle{EX}}^{2}) + v^{2} k^{2})}{(\lambda_{R}^{4} + v^{2} k^{2} (\lambda_{R}^{2} + \lambda_{\scriptscriptstyle{EX}}^{2}))^{3/2}} = - R_{1(3)}^{K,s}.
\end{equation}

\begin{figure*}[t]
	\centering
	\includegraphics[width=0.7\textwidth]{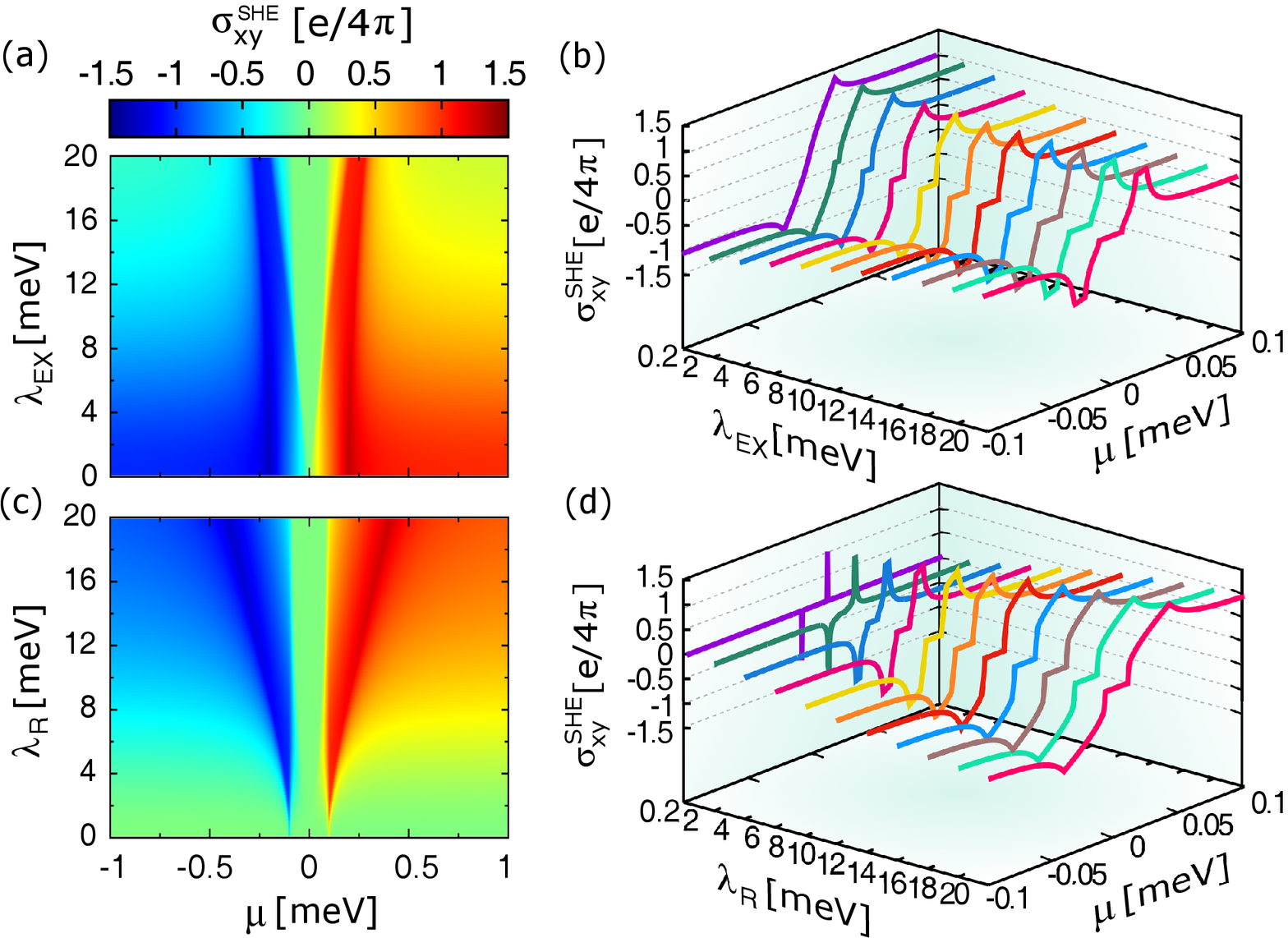}\\
	\caption{(Color online) Spin Hall conductivity in the graphene/YIG system as a function of chemical potential and exchange parameter for fixed Rashba parameter $\lambda_{R} = 10$ meV (a) and as a function of chemical potential and Rashba parameter for fixed exchange parameter $\lambda_{EX} = 10$ meV (b). (c) and (d) represent cross-sections of the density plots in (a) and (b), respectively. }
	\label{SHE}
\vspace{0.3cm}
	\centering
	\includegraphics[width=0.7\textwidth]{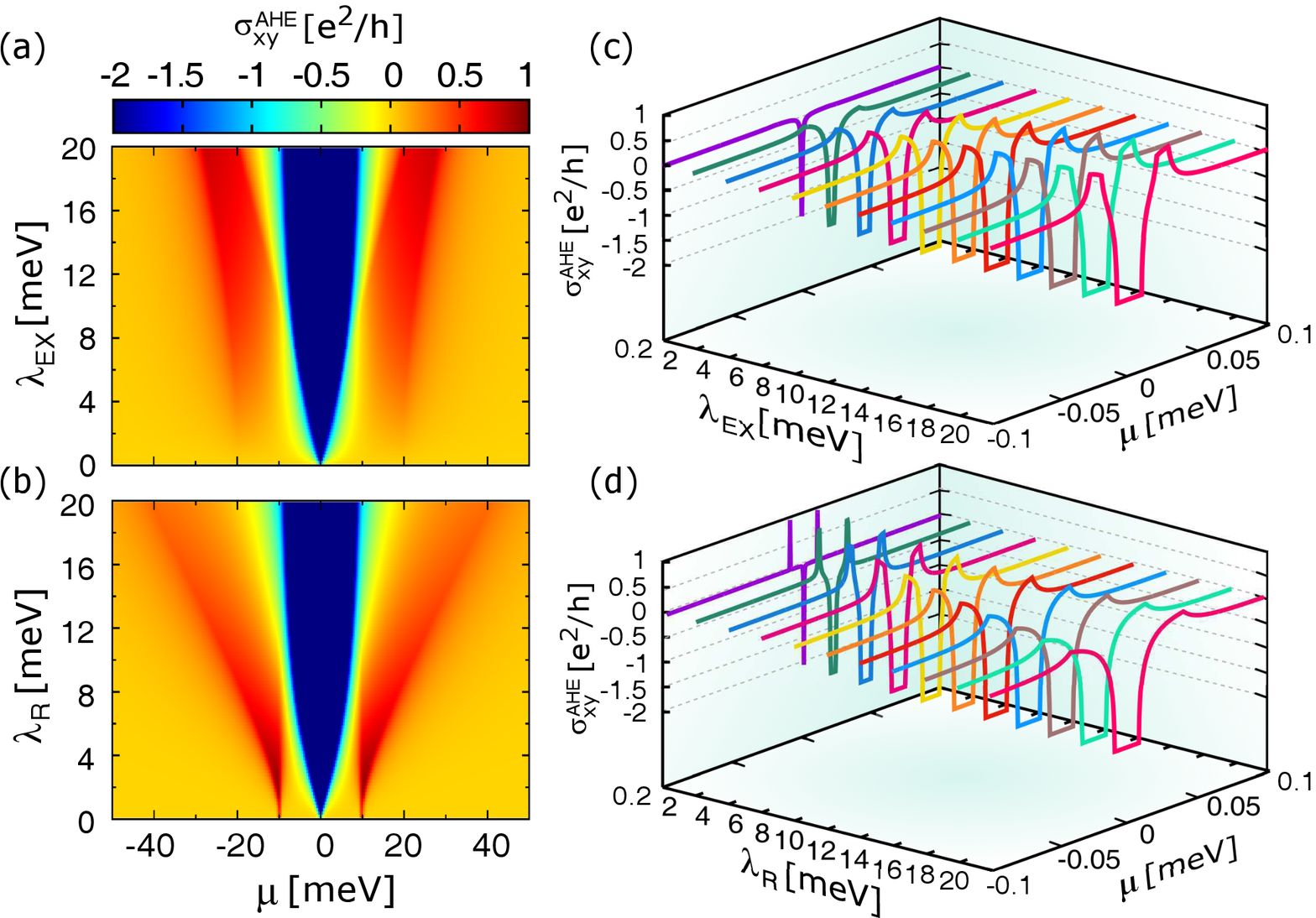}\\
	\caption{(Color online) Anomalous Hall conductivity in the graphene/YIG system as function of chemical potential and exchange parameter  for fixed Rashba parameter $\lambda_{R} = 10$ meV (a) and as a function of chemical potential and Rashba parameter for fixed exchange parameter $\lambda_{EX} = 10$ meV (b). (c) and (d) represent cross-sections of the density plots in (a) and (b), respectively.}
	\label{AHE}
\end{figure*}

Taking the above formulas into account, one can find explicit expressions for the spin Hall conductivity, which are valid in the corresponding  regions of the chemical potential, as described below. These regions can be easily identified when looking at the dispersion curves in Fig.~\ref{DISP_G/YIG}.\\
(i)  $|\mu| > \sqrt{4 \lambda_{R}^{2} + \lambda_{\scriptscriptstyle{EX}}^{2}}$:
\begin{eqnarray}
\sigma_{xy}^{\scriptscriptstyle{SHE}} = \mp \frac{e}{8 \pi} \frac{\lambda_{R}^{2} v^{2}}{\lambda_{R}^{2} + \lambda_{\scriptscriptstyle{EX}}^{2}} \left(\frac{k_{3+}^{2}}{\xi_{3+}} - \frac{k_{3-}^{2}}{\xi_{3-}} \right)\hspace{2cm}\nonumber\\
\pm \frac{e}{4\pi} \frac{\lambda_{R}^{2} \lambda_{\scriptscriptstyle{EX}}^{2}}{(\lambda_{R}^{2} + \lambda_{\scriptscriptstyle{EX}}^{2})^{2}} (2 \lambda_{R}^{2} + \lambda_{\scriptscriptstyle{EX}}^{2}) \left(\frac{1}{\xi_{3+}} - \frac{1}{\xi_{3-}} \right), \hspace{0.2cm}
\end{eqnarray}
with the upper sign for $\mu <0$ and lower for $\mu >0$. \\
(ii) $ \sqrt{4 \lambda_{R}^{2} + \lambda_{\scriptscriptstyle{EX}}^{2}} > |\mu | > \lambda_{\scriptscriptstyle{EX}}$:
\begin{eqnarray}
\sigma_{xy}^{\scriptscriptstyle{SHE}} = \mp \frac{e}{8 \pi} \frac{\lambda_{R}^{2} v^{2}}{\lambda_{R}^{2} + \lambda_{\scriptscriptstyle{EX}}^{2}} \frac{k_{3+}^{2}}{\xi_{3+}}\hspace{2.7cm} \nonumber\\
\pm \frac{e}{4\pi} \lambda_{R}^{2} \lambda_{\scriptscriptstyle{EX}}^{2} \frac{2\lambda_{R}^{2}+ \lambda_{\scriptscriptstyle{EX}}^{2}}{(\lambda_{R}^{2}+ \lambda_{\scriptscriptstyle{EX}}^{2})^{2}} \left( \frac{1}{\xi_{3+}} - \frac{1}{\lambda_{R}^{2}}\right)
\end{eqnarray}
with the upper sign for $\mu <0$ and lower for $\mu >0$. \\
(iii) $  \lambda_{\scriptscriptstyle{EX}} > |\mu | > \sqrt{\frac{\lambda_{R}^{2} \lambda_{\scriptscriptstyle{EX}}^{2}}{\lambda_{R}^{2} + \lambda_{\scriptscriptstyle{EX}}^{2}}}$:
\begin{eqnarray}
\sigma_{xy}^{\scriptscriptstyle{SHE}} = \mp \frac{e}{8 \pi} \frac{\lambda_{R}^{2} v^{2}}{\lambda_{R}^{2} + \lambda_{\scriptscriptstyle{EX}}^{2}} \left(\frac{k_{3+}^{2}}{\xi_{3+}} - \frac{k_{3-}^{2}}{\xi_{3-}} \right)\hspace{2cm}\nonumber\\
\pm \frac{e}{4\pi} \frac{\lambda_{R}^{2} \lambda_{\scriptscriptstyle{EX}}^{2}}{(\lambda_{R}^{2} + \lambda_{\scriptscriptstyle{EX}}^{2})^{2}} (2 \lambda_{R}^{2} + \lambda_{\scriptscriptstyle{EX}}^{2}) \left(\frac{1}{\xi_{3+}} - \frac{1}{\xi_{3-}} \right).\hspace{0.2cm}
\end{eqnarray}
with the upper sign for $\mu <0$ and lower for $\mu >0$. \\
(iv) $- \sqrt{\frac{\lambda_{R}^{2} \lambda_{\scriptscriptstyle{EX}}^{2}}{\lambda_{R}^{2} + \lambda_{\scriptscriptstyle{EX}}^{2}}} < \mu <  \sqrt{\frac{\lambda_{R}^{2} \lambda_{\scriptscriptstyle{EX}}^{2}}{\lambda_{R}^{2} + \lambda_{\scriptscriptstyle{EX}}^{2}}}$ (i.e. in the gap of electronic spectrum):
\begin{eqnarray}
\sigma_{xy}^{\scriptscriptstyle{SHE}} = 0,
\end{eqnarray}
i.e. the spin Hall conductivity vanishes.\\
In the above equations we introduced the  notation:
$k_{3\pm} = \frac{1}{v} \sqrt{\lambda_{\scriptscriptstyle{EX}}^{2} + \mu^{2} \pm 2 \sqrt{\mu^{2} (\lambda_{\scriptscriptstyle{EX}}^{2} + \lambda_{R}^{2}) - \lambda_{\scriptscriptstyle{EX}}^{2} \lambda_{R}^{2}}}$ and $\xi_{3\pm} = \sqrt{\lambda_{R}^{4} + k_{3\pm}^{2} v^{2} (\lambda_{R}^{2} +  \lambda_{\scriptscriptstyle{EX}}^{2})}$

Variation of the spin Hall conductivity with the chemical potential $\mu$ and Rashba and exchange parameters is shown in Fig.~\ref{SHE}. For small values of the exchange parameter, the spin Hall conductivity depends on the chemical potential in a similar way as in graphene on nonmagnetic substrates~\cite{Dyrdal2009}. However, when the exchange coupling increases, the spin Hall conductivity vanishes in the energy gap created by the exchange interaction around $\mu = 0$,  where $\sigma_{xy}^{\scriptscriptstyle{SHE}}=0$. This is clearly seen in Fig.\ref{SHE}b and Fig.\ref{SHE}c, where the platos correspond to the zero spin Hall conductivity in the gap. Vanishing of spin Hall conductivity in the gap is a consequence of the compensation of contributions from the two occupied valence subbands which correspond to opposite spin orientations.
Width of a given plato depends on the strengths of Rashba and exchange couplings. Outside the platos, the absolute value of $\sigma_{xy}^{\scriptscriptstyle{SHE}}$ grows up and upon reaching a maximum decreases with a further increase in $\mu$, tending  to a universal value $e/4\pi$.

\subsection{Anomalous Hall effect}

The anomalous Hall conductivity  can be calculated in a similar way as the spin Hall conductivity. The corresponding formula for the residua, and thus also for the anomalous Hall conductivity, are rather cumbersome, so they are not   presented here. Instead, we show in Fig.~\ref{AHE} only numerical results. First, one can  note that the anomalous Hall conductivity disappears for vanishing Rashba coupling. It also vanishes when the exchange coupling is zero as the system is nonmagnetic. The most interesting feature of the AHE is its quantized value for chemical potentials in the gap formed around $\mu =0$ due to exchange coupling, where  $\sigma_{xy}^{\scriptscriptstyle{AHE}}=-2e^2/h$.
This quantized value is of intrinsic (topological) origin, and is consequence of the fact the Berry curvatures of the bands in the  $K$ and $K'$ points are the same.

\section{Summary}

In this paper we analyzed graphene based hybrid systems, more specifically graphene deposited on magnetic substrates. The key objective was to study the influence of proximity effects, especially of the spin-orbit interaction of Rashba type and the proximity-induced exchange interaction.  Two kinds of systems were considered: (i) graphene deposited on a few atomic monolayers of boron nitride, which in turn was deposited on a magnetic substrate (Co or Ni), and (ii) graphene deposited directly on a magnetic (insulating) substrate like YIG. To describe these systems we assumed the model Hamiltonians which were proposed recently on the basis of  results obtained from {\it ab-initio} calculations.

Our main interest was in the spin, anomalous and valley Hall effects. In addition, we also introduced the valey spin Hall effect. The corresponding conductivities were calculated
in the linear response regime and within the Green function formalism. In the case of graphen/BN/Co(Ni) hybrid system the strength of exchange coupling is controlled by the number of atomic monolayers of BN. Moreover, the atomic structure of BN  leads to a valley gap, which in turn  results in a nonzero valley Hall effect and also in a nonzero valley spin Hall effect. These effects are absent in the case when graphene is deposited directly on YIG. However, anomalous and spin Hall effects can be then observed, with universal quantized values for Fermi level in the energy gap. These universal values follow from topological properties and nonzero Berry curvature.

%\begin{acknowledgments}
\ack
	This work has been supported by the National Science Center in Poland as research
	project No. DEC-2013/10/M/ST3/00488 and by the Polish Ministry of Science and Higher Education (AD)
	through a research project 'Iuventus Plus' in years 2015-2017 (project No. 0083/IP3/2015/73). A.D. also acknowledges support from the Fundation for Polish Science (FNP).
%\end{acknowledgments}

%%%%%%%%%%%%%%%%%%%%%%%
\section*{References}
%%%%%%%%%%%%%%%%%%%%%%%%%%%%%%%%%%%%%%%%%%%%%%%%%%%%%%%%%%%%%%%%%%%%%%%%%%%%%%%%%%%%%%%%%%%%%%%%%%%%%%%%%

\end{document}